\documentclass[usenatbib]{mnras}

\usepackage{graphicx}   
\usepackage{newtxtext,newtxmath}
\usepackage{url}
\usepackage{float}
\usepackage{array}
\usepackage{natbib}
\usepackage{amsmath, amsfonts}

\newcommand{\simgt}{\lower.5ex\hbox{$\; \buildrel > \over \sim \;$}}
\newcommand{\simlt}{\lower.5ex\hbox{$\; \buildrel < \over \sim \;$}}

\title[ORELSE Mass Function]{An Optical Observational Cluster Mass Function at $z\sim1$ with the ORELSE Survey}
\author[Hung et al.]{D. Hung$^{1}$, B.~C. Lemaux$^{2}$, R.~R. Gal$^{1}$, A.~R. Tomczak$^{2}$, L.~M. Lubin$^{2}$, O. Cucciati$^{3}$, 
\newauthor D. Pelliccia$^{2,4}$, L. Shen$^{2}$, O. Le F\`{e}vre$^{5}$, G. Zamorani$^{3}$, P-F. Wu$^{6}$, D.~D. Kocevski$^{7}$,
\newauthor C.~D. Fassnacht$^{2}$, G.~K. Squires$^{8}$ \\
$^{1}$ University of Hawai'i, Institute for Astronomy, 2680 Woodlawn Drive, Honolulu, HI 96822, USA \\
$^{2}$ Department of Physics \& Astronomy, University of California, Davis, One Shields Ave., Davis, CA 95616, USA \\
$^{3}$ INAF - Osservatorio di Astrofisica e Scienza dello Spazio diBologna, via Gobetti 93/3 - 40129 Bologna - Italy \\
$^{4}$ UCO/Lick Observatory, Department of Astronomy \& Astrophysics, UCSC, 1156 High Street, Santa Cruz, CA, 95064, USA \\
$^{5}$ Aix-Marseille Universit{\'e}, CNRS, LAM (Laboratoire d'Astrophysique de Marseille) UMR 7326, 13388 Marseille, France \\
$^{6}$ National Astronomical Observatory of Japan, Osawa 2-21-1, Mitaka, Tokyo 181-8588, Japan \\
$^{7}$ Department of Physics and Astronomy, Colby College, Waterville, ME 04961, USA \\
$^{8}$ Spitzer Science Center, California Institute of Technology, M/S 220-6, 1200 E. California Blvd., Pasadena, CA 91125, USA}

\date{Accepted XXX. Received YYY; in original form ZZZ}

\pubyear{2020}

\begin{document}
\label{firstpage}
\pagerange{\pageref{firstpage}--\pageref{lastpage}}
\maketitle

\begin{abstract}
We present a new mass function of galaxy clusters and groups using optical/near-infrared wavelength spectroscopic and photometric data from the Observations of Redshift Evolution in Large-Scale Environments (ORELSE) survey. At $z\sim1$, cluster mass function studies are rare regardless of wavelength and have never been attempted from an optical/near-infrared perspective. This work serves as a proof of concept that $z\sim1$ cluster mass functions are achievable without supplemental X-ray or Sunyaev-Zel'dovich (SZ) data. Measurements of the cluster mass function provide important contraints on cosmological parameters and are complementary to other probes. With ORELSE, a new cluster finding technique based on Voronoi tessellation Monte-Carlo (VMC) mapping, and rigorous purity and completeness testing, we have obtained $\sim$240 galaxy overdensity candidates in the redshift range $0.55<z<1.37$ at a mass range of $13.6<\log(M/M_{\odot})<14.8$. This mass range is comparable to existing optical cluster mass function studies for the local universe. Our candidate numbers vary based on the choice of multiple input parameters related to detection and characterization in our cluster finding algorithm, which we incorporated into the mass function analysis through a Monte-Carlo scheme. We find cosmological constraints on the matter density, $\Omega_{m}$, and the amplitude of fluctuations, $\sigma_{8}$, of $\Omega_{m} = 0.250^{+0.104}_{-0.099}$ and $\sigma_{8} = 1.150^{+0.260}_{-0.163}$. While our $\Omega_{m}$ value is close to concordance, our $\sigma_{8}$ value is $\sim2\sigma$ higher because of the inflated observed number densities compared to theoretical mass function models owing to how our survey targeted overdense regions. With Euclid and several other large, unbiased optical surveys on the horizon, VMC mapping will enable optical/NIR cluster cosmology at redshifts much higher than what has been possible before.
\end{abstract}

\section{Introduction}

Cosmological models seek in part to explain the growth and distribution of large-scale structure in the universe. One such quantifying metric is the cluster mass function, which describes the number density of galaxy clusters as a function of their mass. How the mass function evolves over time will depend on cosmological parameters, and thus measuring the mass function over wide redshift ranges offers the power of greater statistical leverage \citep[see, e.g.,][]{allen11}. Different cosmologies in theoretical mass functions show non-negligible discrepancies in the predicted number counts of clusters \citep[e.g.,][]{pacaud18}, motivating the need for comparisons with observational data. 

Constraints on cosmological parameters can be obtained through fitting a number of independent probes, such as the cosmic microwave background (CMB) anisotropy \citep[e.g.,][]{planck, wmap}, the brightness/redshift relation for type Ia supernovae \citep[SNe; e.g.,][]{riess98, perlmutter99}, and baryon acoustic oscillations (BAO) data \citep[e.g.,][]{eisenstein05}. The cluster mass function can be used to constrain the matter density, $\Omega_{m}$, and the amplitude of fluctuations on the scale of 8 $h^{-1}$ Mpc, $\sigma_{8}$, by fitting the predicted halo abundance, the halo mass function. $\sigma_{8}$ shows a strong degeneracy with $\Omega_{m}$ when determined from cluster abundance data. However, the confidence levels of the $\Omega_{m}$-$\sigma_{8}$ likelihoods found by the cluster mass function are advantageously almost orthogonal to those found by the CMB \citep[e.g.,][]{rozo10}. Combining the two probes therefore helps break the degeneracy between $\Omega_{m}$ and $\sigma_{8}$ and reduce their uncertainties, while BAO and SNe studies can constrain $\Omega_{m}$ independent of $\sigma_{8}$ \citep[e.g.,][]{vikhlinin09, abdullah20}.

The first analytical expression of the halo mass function was derived by \citet{press74}, followed by \citet{bond91, lee98, sheth01}. However, the rise of N-body simulations helped reveal limitations in the existing models, and the most recent halo mass functions have been calibrated using numerical results. These models are chiefly distinguished between two widely used halo definitions. Haloes may be defined using the spherical overdensity \citep[SO;][]{lacey94} algorithm, where spherical apertures are placed around isolated density peaks, such that the mean interior density is some set multiple relative to the background or critical density. Haloes may also be defined with the Friends-of-Friends \citep[FoF;][]{davis85} algorithm, where a particle is matched with neighbors within a given linking length, and those neighbors are matched with other neighbors until no more are found. The final group of particles then represents an isodensity contour in space. SO and FoF masses are strongly correlated for relaxed, isolated haloes \citep{white01, tinker08}, but irregular haloes can cause significant disagreement. Most theoretical models have followed the convention of \citet{jenkins01} and used FoF haloes in order to obtain a more universal halo mass function independent of redshift or cosmology. However, SO haloes tend to be preferred for comparisons to observational studies, due to the more direct link with how virialized structures are defined in spherical apertures.

Determining a cluster mass function from observational data requires a cluster sample where cluster masses have been estimated either directly (as in, e.g., weak gravitational lensing) or by using an observational proxy. Typically, the cluster sample is X-ray selected, and the masses are derived through more indirect proxies such as X-ray luminosity or optical cluster richness \citep[e.g.,][]{reiprich02, mantz08, vikhlinin09, wen10, pacaud18, costanzi19}. However, the resulting cluster mass function can have large uncertainties due to factors such as the scatter in the mass scaling relations as well as incompleteness in the cluster sample due to selection biases or other observational effects. These issues are especially a concern at redshift $z\simgt0.5$, where the intracluster medium (ICM) begins to be underdeveloped, particularly for intermediate or low-mass clusters. Many clusters have been found to be X-ray underluminous, compared to what was suggested by their dynamics \citep[e.g.,][]{rumbaugh18} or the luminosity-mass relation with weak lensing masses \citep[e.g.,][]{giles15}. Cluster samples selected by other means often see a sizable portion with no detected X-ray counterpart at $z\sim1$ \citep[e.g.,][]{popesso07, rumbaugh18}.

Cluster mass function studies have often supplemented their X-ray selected samples with other observations in order to obtain more reliable mass estimates, such as weak lensing masses \citep[e.g.,][]{dahle06} or virial masses from large-scale redshift surveys \citep[e.g.,][]{rines07}. Clusters may be found through searching for signatures of the thermal Sunyaev-Zel'dovich \citep[SZ;][]{sz72} effect \citep[e.g.,][]{staniszewski09, menanteau10, bocquet19, bleem20, huang20, hilton20}, which is a distortion in the CMB blackbody spectrum as a result of Compton scattering of CMB photons by the hot ICM. The SZ effect is unaffected by the clusters' surface brightness dimming and is thus insensitive to redshift. SZ surveys are thus able to detect clusters at all redshifts above a certain mass threshold set by the detection limits of the SZ signal \citep{birkinshaw99, carlstrom02}. Because of the high sensitivity needed, even recent studies \citep[e.g.,][]{bocquet19,bleem20, huang20, hilton20} have been limited to cluster samples with masses greater than $10^{14} M_{\odot}$. In contrast, X-ray and optical/near-infrared (NIR) surveys are more effective at finding low-mass clusters, particularly at lower redshifts.

More recently, the growing scale of photometric and spectroscopic surveys at optical and NIR wavelengths have enabled cluster searches independent of any X-ray data. Such searches identify clusters by using galaxies to trace mass overdensities \citep[e.g.,][]{abell58, oke98, rykoff16}. Though there have been several successful cluster searches done at optical wavelengths, cluster mass function studies at optical and NIR wavelengths have been scarce and so far limited to the local universe. Such studies use optically selected catalogs with masses derived from supplemental weak lensing or X-ray data \citep[e.g.,][]{rozo10, costanzi19, kirby19} or through the virial mass theorem \citep[e.g.,][]{abdullah20}. Beyond constraining cosmology, contrasting cluster mass functions with X-ray and optically selected samples at different redshifts could yield key insights on structure formation and development of the ICM. While at least some attempts have been made at X-ray and SZ wavelengths at redshifts up to $z\sim1-1.5$ to bridge the gap between theory and observation \citep[e.g.,][]{pacaud18, bocquet19}, the same cannot be said for optical studies. 

Our work in this paper aims to similarly derive a cluster mass function\footnote{Our sample includes overdensity candidates with masses as small as $\sim10^{13.5} M_{\odot}$, which fall below the typically defined mass limits of galaxy clusters and instead would traditionally be regarded as groups. Though these structures are all included in the mass function, we use the term ``cluster mass function'' in this paper for the sake of brevity.} from an optical/NIR perspective for the first time outside of the relatively local universe. In \citet{hung20}, we found galaxy clusters using a powerful new technique known as Voronoi tessellation Monte-Carlo (VMC) mapping and apply it to optical and NIR photometric and spectroscopic data over the redshift range $0.55<z<1.37$. Unlike other cluster search algorithms, VMC mapping makes no assumptions about cluster geometry or morphology. With VMC mapping, we count all galaxies irrespective of color to a limit of stellar masses $\simgt10^{10} M_{\odot}$ to trace overdensities, independent of the ICM emission. 

In searches of clusters with X-ray observations, there is a possibility of observing a decreasing number of systems at a given X-ray luminosity with increasing redshift. In such a case, an ambiguity would exist in the interpretation of the trend as this behavior could either be attributed to intermediate- to high-redshift structures of a given mass having an underdeveloped ICM relative to the local counterparts \citep[as is true for at least some ORELSE systems, see][]{rumbaugh18}, a true lack of structure at higher redshift, or some combination of the two. The same ambiguity does not exist in optical/NIR cluster searches with spectroscopically confirmed redshifts as galaxies will presumably always trace clusters. In our search, because we indiscriminately count galaxies without constraining ourselves to any particular subpopulations such as the red sequence, we should be able to detect a cluster so long as it is galaxy-rich with any type of galaxies.

In \citet{hung20}, we demonstrated VMC mapping's sensitivity to detecting unprecedentedly low mass structures, quantifying purity and completeness estimates down to total masses of $10^{13.5} M_{\odot}$. Our search recovered 51 previously known structures and found 402 new overdensity candidates, with estimated masses between 10.2 $<\log(M/M_{\odot})<$ 14.8\footnote{As we can only correct for purity and completeness  down to masses of $10^{13.5} M_{\odot}$, it is possible that many of the overdensity candidates with smaller masses are spurious detections. We refer the reader to \S6.1.1 in \citet{hung20} for a more in-depth discussion.} and a spectroscopic redshift fraction of at least 5\%. In this paper, we seek to derive a cluster mass function drawn from this sample. 

This paper is organized as follows: In \S\ref{sec.data}, we briefly review the photometric and spectroscopic data we used and our overdensity candidate detection method. In \S\ref{sec.finding}, we go over several parameters that affect the overdensity candidate sample. In \S\ref{sec.massfuncdata}, we describe how we transform these parameters and their varying overdensity candidate samples into one mass function. In \S\ref{sec.massfunctheory}, we compare our observational mass function with a theoretical model to fit for $\Omega_{m}$ and $\sigma_{8}$. In \S\ref{sec.discuss}, we discuss the implications of our findings as well as a few other cluster mass function studies and highlight where our methodolgy could be useful with data from future surveys. Finally, we present a summary of this work in \S\ref{sec.conclusion}. Unless otherwise noted, we use a flat $\Lambda$CDM cosmology throughout this paper, with $H_{0}$ = 70 km s$^{-1}$ Mpc$^{-1}$, $\Omega_{m}$ = 0.27, and $\Omega_{\Lambda}$ = 0.73. All reported distances are in proper units.

\begin{figure*}
\centering
\includegraphics[width=2\columnwidth]{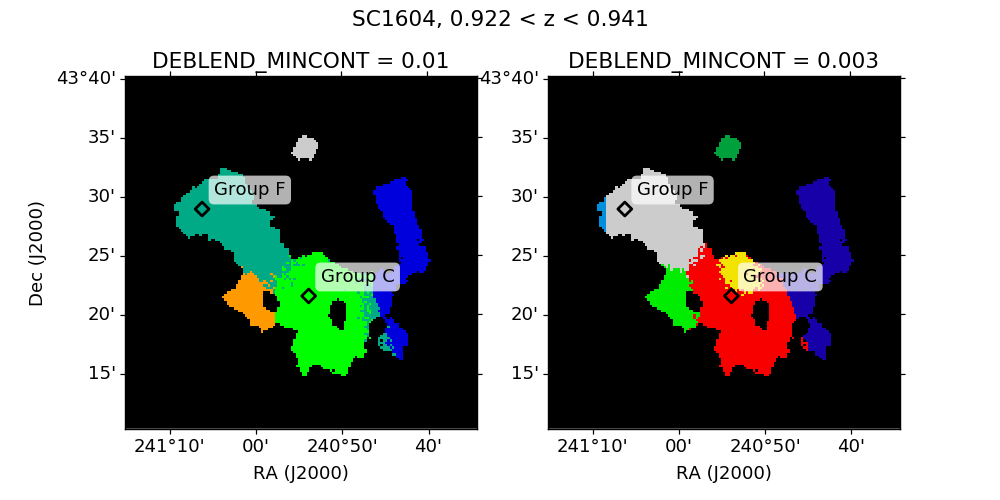}
\caption{Example SExtractor-generated segmentation maps of a blended structure in the SC1604 supercluster at $z\sim0.9$ located in one of the ORELSE fields. The two panels show the segmentation map generated from two different values of the minimum contrast DEBLEND\textunderscore MINCONT. Two sub-components of the SC1604 supercluster which are spectroscopically confirmed at this redshift, groups C and F, have their fiducial positions denoted in the maps. The colors here are arbitrarily assigned; each color simply represents a single detection in SExtractor. Note that SExtractor's segmentation maps often have display glitches, such as group F's disparate pixels in the left panel and the sharp vertical boundary in the right panel. Only unique colors indictate what SExtractor considers as separate substructure. As the DEBLEND\textunderscore MINCONT decreases from the left to right panel, the number of substructures increases from five to seven.}
\label{fig.deblends}
\end{figure*}

\section{Data}\label{sec.data}

Our previous work in \citet{hung20} searched for serendipitous cluster candidates in the Observations of Redshift Evolution in Large-Scale Environments \citep[ORELSE;][]{lubin09} survey, a large multi-wavelength photometric and spectroscopic campaign targeted at several known large-scale structures over redshifts of $0.6<z<1.3$. It was designed to look for surrounding large-scale structure in each field, but it also probed the full dynamic range of environments at all redshifts by targeting galaxies along the line-of-sight \citep{gal08, lubin09}. Over 15 fields, ORELSE has a combined $\sim5$ square degrees of deep imaging and a projected spectroscopic footprint of $\sim1.4$ degrees. The optical ($BVriz$) imaging typically ranged from depths of $m_{AB}$ = 26.4 in the $B$-band to $m_{AB}$ = 24.6 in the $z$-bands. The NIR ($JK$, {\it Spitzer}/IRAC) imaging reached typical depths of $m_{AB}$ = 21.9 and 21.7 respectively in the $J$ and $K$/$K_{s}$ bands \citep{tomczak19}. Its unprecedented spectroscopic coverage includes $\sim11,000$ high quality spectroscopic objects and spectroscopic completeness of 25\% to 80\% among known structures \citep{lemaux19}. Additionally, the spectral member population has been found to be broadly representative of the underlying galaxy population \citep{shen17, lemaux19}. ORELSE's extensive dataset provides thousands of high-quality photometric and spectroscopic redshifts ideal for a cluster search.

We identified galactic overdensities using a powerful new technique, Voronoi tessellation Monte-Carlo (VMC) mapping, described in detail in \citet{lemaux18} and applied to look specifically for structure in ORELSE in \citet{hung20}. A Voronoi tessellation is a density field estimator that splits a 2D plane by assigning a polygonal cell to every object in the plane whose area is the region closer to its host object than any other object. The cell size is thus inversely proportional to the density at a given location. For each ORELSE field, we separate our galaxy catalogs into redshift slices of approximately $\pm$1500 km s$^{-1}$ in velocity space and apply the tessellation to each slice. The redshift slices are defined such that neighboring slices have 90\% overlap to minimize chances of splitting individual structures across slices. 

For each slice, we have galaxies with spectroscopic redshifts, $z_{spec}$, and galaxies with photometric redshifts, $z_{phot}$. The photometric redshifts have much higher uncertainties than the spectroscopic redshifts, which we account for with our VMC technique. For each Monte-Carlo realization of a slice, we  Gaussian sample the PDF of each galaxy's $z_{phot}$. As a result, some galaxies fall in or out of the redshift boundaries of the slice. We then perform the Voronoi tessellation on all the $z_{spec}$ and $z_{phot}$ galaxies in the slice. We repeat this 100 times, and the final VMC map of the slice is then computed by median combining the densities from all realizations. For full details on the VMC methodology within the context of ORELSE, see \citet{hung20}. Overdensities are first found in the redshift slices (see \S\ref{sec.detect}), and then linked together across neighboring slices (see \S\ref{sec.linking}).

\begin{figure}
\includegraphics[width=\columnwidth]{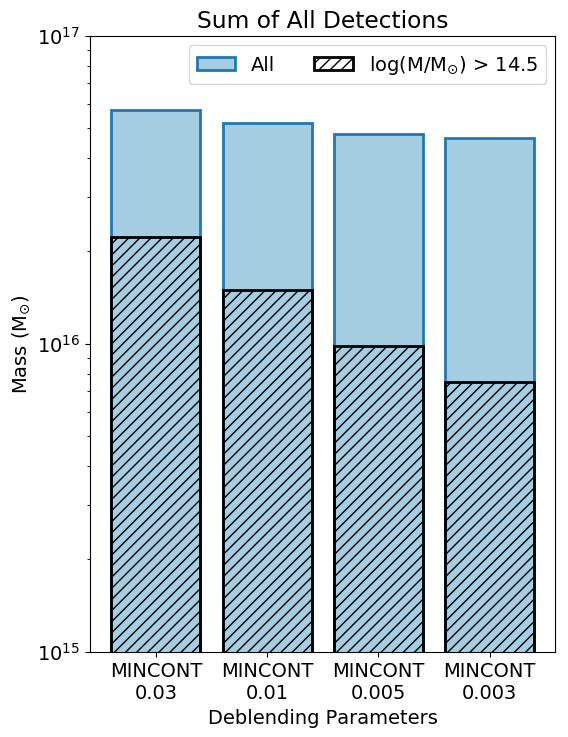}
\caption{We plot the total mass of all overdensity candidates we find while varying the SExtractor deblending parameter DEBLEND\textunderscore MINCONT. The deblending becomes finer for smaller values of DEBLEND\textunderscore MINCONT. Following our expectations, finer deblending yields more low-mass overdensity candidates as more high-mass structures are broken up. Between the DEBLEND\textunderscore MINCONT extremes plotted, the total mass decreases by 19\%, while the fraction of high mass candidates drops by 66\%.}
\label{fig.allmass}
\end{figure}

\section{Finding Structure}\label{sec.finding}

How we find and catalog galaxy overdensity candidates depends on several independent parameters, ranging from how large an overdensity must be for detection to peculiarities on how we translate the overdensity we observe to a total mass. In \citet{hung20}, our goal was to establish VMC mapping as a viable tool for finding overdensities. We thus adopted a set of parameters best suited for the general case of detecting any structure at all and left the specifics of fine-tuning the resulting overdensity candidate sample to future work. We revisit our parameters in this work as we now require crucial informaiton such as the proper number of overdensities at each mass threshold in order to build the cluster mass function. In this section, we describe the effects of each revelant parameter, and in \S\ref{sec.massfuncdata}, we go over which values we use for our cluster mass function. We encourage the reader to refer to \citet{hung20} where these parameters are described in greater detail.

\subsection{Detection and Deblending in SExtractor}\label{sec.detect}
We search for significant overdense regions in our VMC maps using the standard photometry software package Source Extractor \citep[SExtractor;][]{bertin96}. SExtractor's DETECT\textunderscore THRESH parameter sets how much higher the density floor must be for a valid detection relative to the RMS noise in the background. For example, a 4$\sigma$ DETECT\textunderscore THRESH stipulates that detections must be at least four times the background RMS. In \citet{hung20}, we found that DETECT\textunderscore THRESH values of 4 and 5$\sigma$ performed similarly in terms of purity and completeness, but we decided to use 4$\sigma$ to maximize our chances of detecting smaller overdensity candidates in that work.

Often, galaxy clusters are located in close proximity to each other, and show up in SExtractor as single detections. Deblending in SExtractor refers to separating these detections out to their subcomponents so that we can identify the individual clusters. SExtractor has two parameters related to deblending: the number of deblending sub-thresholds DEBLEND\textunderscore NTHRESH and the minimum contrast DEBLEND\textunderscore MINCONT. The deblending sub-thresholds refer to the number of exponentially spaced levels from the detection floor to the peak of the detection. Substructure is identified with the minimum contrast DEBLEND\textunderscore MINCONT parameter, which is how large the overdensity in a substructure must be compared to the total overdensity in the entire structure to be counted as a separate detection. Of the two deblending parameters, we choose to focus on the DEBLEND\textunderscore MINCONT parameter as it is more sensitive to change. As DEBLEND\textunderscore MINCONT decreases, the more SExtractor splits apart a single structure (Fig. \ref{fig.deblends}).

Previously in \citet{hung20}, we elected to adopt a DEBLEND\textunderscore NTHRESH of 32 and DEBLEND\textunderscore MINCONT of 0.01. We deemed these parameters as acceptable as they were able to separate some known structures while also avoiding splitting others up. However, not every blended grouping of known structures was able to be separated with these deblending parameters. For an unbiased cluster mass function, we must be able to properly separate larger conglomerates of structure by way of carefully choosing the optimal set of deblending parameters. Not doing so would lead to an overabundance of high mass overdensity candidates and a depletion of low mass overdensity candidates.

We have a measure of each overdensity candidate's mass by way of its isophotal flux $F$, a measurement of density calculated by SExtractor, of the form:

\begin{equation}\label{eq.massfit}
\log(M / M_{\odot}) = a + b F^{c} e^{-(F/d)}
\end{equation}

where $a$, $b$, $c$, and $d$ are scalar constants. We fit this quantity with the virial masses of the previously known structures in ORELSE to obtain a general flux to mass relation. The mass zero point was calibrated with the virial masses of the most spectroscopically well-studied clusters and groups in ORELSE, which generally had spectral fractions of $>50$\% and an average of 24 spectroscopic members per structure. The virial masses have been found to be comparable within the error bars of independent X-ray, lensing, and SZ measurements where available \citep[see e.g.,][]{clowe98, margoniner04, valtchanov04, jee06, maughan06, muchovej07, rzepecki07, stott10, piffaretti11, lagana13, pratt20} as well as statistically consistent with the masses we estimate from the overdensity maps directly using the method described in \citet{cucciati18}. We refer the reader to \S6.1 of \citet{hung20} for further discussion of our mass calibration as well as comparisons with other mass estimation methods.

Previously in \citet{hung20}, we found best-fit values of $a = 15.691 \pm 0.010$, $b = -2.641 \pm 0.033$, $c = -0.327 \pm 0.039$, and $d = 124.174 \pm 0.740$ for equation \ref{eq.massfit}. The exact mass fit will vary in this work as the detection floor set by the choice of DETECT\textunderscore MINCONT will significantly change the total isophotal flux values, but we find negligible differences with respect to the choice in deblending parameters. The overall mass, summed over all overdensity candidates we find, remains relatively unchanged as we drop the DETECT\textunderscore MINCONT parameter, while the fraction of high mass candidates significantly drops (Fig. \ref{fig.allmass}).

\begin{figure*}
\centering
\includegraphics[width=2\columnwidth]{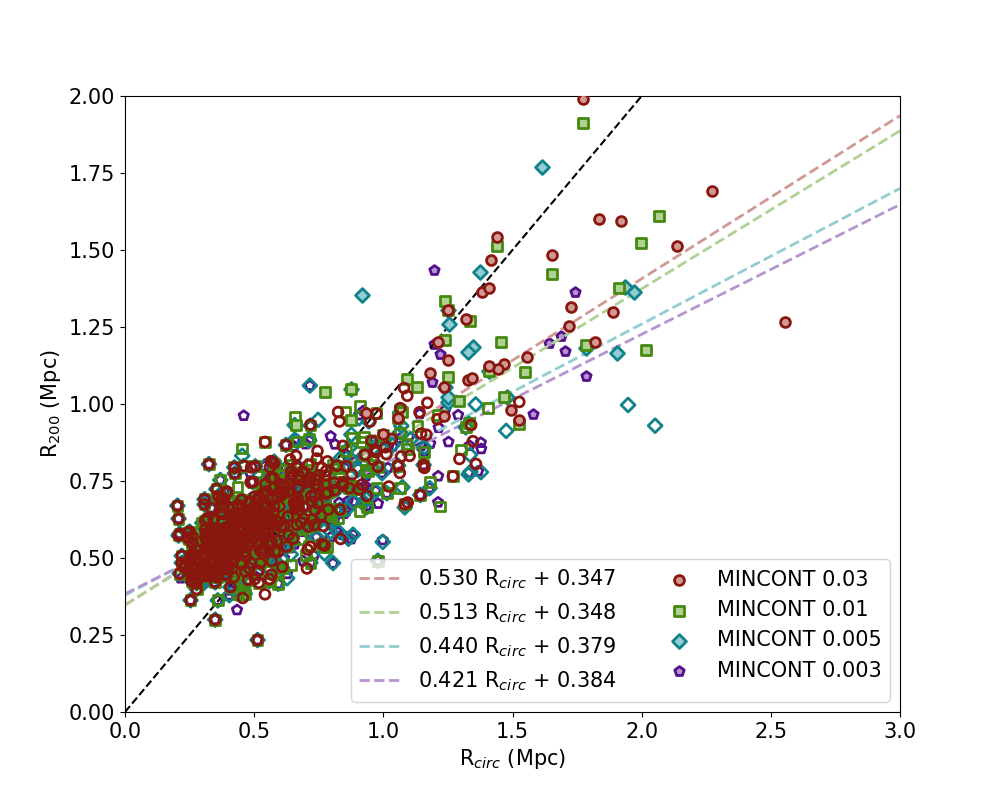}
\caption{In an attempt to better match our cluster mass function to theoretical halo mass function models, we can scale our overdensity candidates circular radii $R_{\mathrm{circ}}$ to their equivalent $R_{200}$ radii, assuming their mass is equal to $M_{200}$. We plot linear fits of $R_{\mathrm{circ}}$ and $R_{200}$ for our four different DEBLEND\textunderscore MINCONT parameters using a fixed DETECT\textunderscore THRESH of 4$\sigma$. We calculate $R_{200}$ according to Equation \ref{eq.r200}. We measure $R_{\mathrm{circ}}$ from the largest SExtractor detection by area in each overdensity candidate. The scatter points represent all found overdensity candidates for each DEBLEND\textunderscore MINCONT parameter. The filled scatter points represent the overdensity candidates with masses $\log(M_{\mathrm{tot}}/M_{\odot}) > 14.5$. The fits between the DEBLEND\textunderscore MINCONT parameters do not significantly change. $R_{\mathrm{circ}}$ and $R_{200}$ appear to be closest to equal (the dashed black line) below around 0.6 Mpc. }
\label{fig.r200_rcirc}
\end{figure*}

\subsection{Linking into Candidates}\label{sec.linking}
SExtractor finds individual overdensities in each redshift slice of a VMC map. We link these overdensities over successive redshift slices in order to obtain a single overdensity candidate. We start with a given SExtractor detection in a redshift slice. Then, we look in the neighboring redshift for any detections with a barycenter within an RA-DEC distance equal to or less than the linking radius we set. The smaller the linking radius, the closer the detections must be in order to be linked together. If we find a match, we take the isophotal flux-weighted barycenter between the two detections and continue our search into the next redshift slice. If there are multiple matches, we take both as separate linked chains and continue the search until no more matches are found. Once we complete the search for every detection in a redshift slice, we move on to the next redshift slice and repeat the same search. For each linked chain of at least five SExtractor detections, we apply a Gaussian fit of the isophotal fluxes of the detections and accept the chain as an overdensity candidate if the Gaussian fit converges.

Because we link every detection to every possible match, we have the same detections assigned to more than one linked chain. Thus, we would overcount the number of overdensity candidates we had if we treated each chain as a new unique candidate. In \citet{hung20}, we resolved this issue by sorting the overdensity candidates by greatest Gaussian fit amplitude and removing any other candidates that fell within 0.7 Mpc and $\Delta z < 0.2$ of their centroids. Though this crude removal process likely left in a few duplicates and eliminated some valid candidates, our work in \citet{hung20} was primarily concerned with establishing the VMC technique for finding overdensities for the general case rather than precise optimization for our particular set of fields.

For the purpose of constructing a cluster mass function, however, including the smaller substructures we originally eliminated is paramount. For this work, we revised our linking scheme by employing a goodness of fit test in order to remove only the duplicate detections. We first remove all linked chains that are complete subsets of other larger chains. Then, we apply our Gaussian fit for each linked chain. We measure the goodness of fit with the coefficient of determination, $R^2$. The $R^2$ statistic ranges between 0 and 1, with the latter indicating a perfect agreement between the model and data. We sort our linked chains by their $R^2$ values from high to low. We accept the first linked chain as an overdensity candidate, and we remove all other linked chains that include any of the same SExtractor detections as the accepted candidate. We repeat this iteratively with the next highest $R^2$ linked chain until no SExtractor detections are shared between any of the candidates. We emphasize that this removal process only eliminates duplicate detections of the same overdensity candidates from our catalog; no real structure or substructure is lost as a result of this process.

Ignoring the linked chains that were complete subsets of another, the removal process eliminated as few as 20 to over 1000 linked chains across all ORELSE fields depending on how sensitive we set the detection and deblending parameters in \S\ref{sec.detect}. Despite this wide range in removals, the number of linked chains remaining was fairly robust, typically being between a total of 300 to 400, so we consistently have around the same number of overdensity candidates after removing all duplicate detections. More linked chains are removed by number at lower redshifts due to a greater abundance of detections, though the percentage of removals is not sensitive to redshift. We note that this methodolgy will disfavor irregularly shaped structures where the velocity distribution deviates appreciably from a Gaussian, though they are likely still picked up in many cases as our goodness of fit test is a relative measure.

\subsection{Exclusion of Previously Known Structures}\label{sec.prev}
ORELSE was designed to target massive known clusters. Thanks to the high levels of spectroscopy around these systems, we found a few dozen more clusters and groups nearby in the fields on an initial, primarily spectroscopic search. In \citet{hung20}, we optimized our choice of SExtractor parameters in part based on how well we could recover all known structures in the ORELSE fields, both those the survey was targeted at and those found with spectroscopy. The inclusion of all these known structures will bias a mass function high relative to structures found in a field survey. This is particularly true at the high mass end as the known structures are among the highest mass overdensities we detect with our technique. In addition, there were a small number of structures, such as clusters B and C in RXJ1716 \citep[see \S4.3.1 in][]{hung20}, that we were not able to separate no matter how fine we set the deblending parameters. We would pick up these blended structures as single overdensities and thus overestimate their masses. By excluding the previously known structures we recover from our cluster mass function calculations, we can possibly avoid biasing our data towards higher mass overdensities. Regardless of this removal, it is likely that the mass function will still be biased high because of additional structures around the targeted structures that were missed by the original spectral search.

\subsection{Correcting to $R_{200}$}
Cluster mass function studies typically compare to theoretical models that calculate the dark matter halo mass function. Dark matter haloes are typically defined within spherical apertures of radii $R_{200}$ corresponding to an overdensity $\Delta = 200$. As we cannot measure haloes from our observational data, we look to scaling the effective circular radius, $R_{\mathrm{circ}}$, of each overdensity candidate to their $R_{200}$ radii. Each overdensity candidate is made up of a series of SExtractor detections we linked together over several redshift slices. We obtain $R_{\mathrm{circ}}$ by taking the largest SExtractor detection by isophotal area in an overdensity candidate and finding its effective circularized radius. We derive $R_{200}$ by treating the overdensity candidate's estimated mass from equation \ref{eq.massfit} in \S\ref{sec.detect} as equal to $M_{200}$ with:

\begin{equation}\label{eq.r200}
R_{200} =  \left ( \frac{G M_{200}}{100 H^2(t)} \right )^{1/3}
\end{equation}

where $H(t)$ is the Hubble parameter and $G$ is the gravitational constant. 

In Figure \ref{fig.r200_rcirc}, we plot $R_{\mathrm{circ}}$ and $R_{200}$ values for four different DEBLEND\textunderscore MINCONT parameters using a DETECT\textunderscore THRESH of 4$\sigma$. Above approximately 0.6 Mpc, $R_{\mathrm{circ}}$ predominantly outpaces $R_{200}$. In other words, it is likely we are estimating a mass for larger overdensity candidates at an effective radius larger than $R_{200}$. The disagreement between the two radius measures implies that our assumption that our mass estimate is equal to $M_{200}$ is incorrect, which means the comparisons between our observed mass function and the theoretical mass functions may also be off due to the latter using $R_{200}$\footnote{For transparency, we note that we also allow the theoretical value to vary to account for the imprecision in this process. More details can be found in \S\ref{sec.massfunctheory}.}. This indicates a possible need to scale down the masses of such overdensity candidates to the mass enclosed by their $R_{200}$ radii to match the comparisons we make with the theoretical halo mass function in \S\ref{sec.massfunctheory}.

Equation 37 of \citet{coe10} gives the mass of a Navarro-Frenk-White (NFW) dark matter halo within a sphere of radius $r = x r_s$ as:

\begin{equation}
M(r) = 4 \pi \rho_s r_s^3 \left ( \ln (1 + x) - \frac{x}{1+x}\right )
\end{equation}

where $\rho_s$ is the scale density, $r_s$ is the scale radius, and $x$ is a multiplicative factor. We use this equation to calculate the quotient of the mass enclosed at $R_{200}$, $M_{200}$, and the mass enclosed at some generalized radius. From equation 1 of \citet{coe10}, the scale radius is equivalent to $r_s = (C_{vir}/r_{vir})^{-1}$, where $C_{vir}$ is the concentration at the virial radius, and $r_{vir}$ is the virial radius. We assume that $r_{vir} = R_{200}/1.14$ and estimate that $C_{vir} \approx 3.5$ for our higher mass overdensity candidates \citep{duffy08}. Taking the ratio $\kappa$ of the masses enclosed in $R_{\mathrm{circ}}$ over $R_{200}$ reduces to:

\begin{equation}
\kappa =  \frac{\ln(5) - 4/5}{\left (\ln(1+4x)-\frac{4x}{1+4x} \right )}
\end{equation}

\begin{equation}
x = \frac{R_{\mathrm{circ}}}{R_{200}} = \frac{(1 - b/R_{200})}{m}
\end{equation}

where $m$ and $b$ are the slope and intercept of the $R_{\mathrm{circ}}$ and $R_{200}$ linear fit.

For a given overdensity candidate, we would multiply its mass by $\kappa$ to scale it back to $M_{200}$. As overdensity candidates with $R_{200} \leq 0.6$ Mpc not only have very low flux values but also $R_{200} \approx R_{\mathrm{circ}}$, we consider this correction only for candidates with $R_{200} > 0.6$ Mpc. $\kappa$ decreases with mass, giving typical values of 0.90, 0.71, and 0.64 for masses of $10^{14} M_{\odot}$, $10^{14.5} M_{\odot}$, and $10^{15} M_{\odot}$ respectively.

\section{Building the Cluster Mass Function}\label{sec.massfuncdata}

We represent the cluster mass function as a cumulative distribution, where we plot the number density $N(>M)$ for a given mass. In \citet{hung20}, we constructed several mock candidate catalogs to estimate our purity and completeness numbers. The mock catalogs sampled slightly different mass ranges depending on the redshift, but they inclusively covered $\log (M/M_{\odot})$ = 13.64 to 14.81. We use 10 equally logarithmically spaced points in the same mass range for our mass function. The spacing between the mass bins do not affect the measured number densities as long as they are wider than the average mass uncertainties. Because we have purity and completeness estimates, and associated uncertainties on those purity and completeness values, as functions of redshift, mass, and spectroscopic function, we do not need to rely on using a $V_{max}$ method to limit our sample to where we have high completeness.

\subsection{Mass Bin Assignment}\label{sec.massbin}
We assign the overdensity candidates to the mass bins using a Monte-Carlo method. For each overdensity candidate, we have an estimate of its mass and redshift and their associated uncertainties. We Gaussian sample each to obtain a new mass $M_i$ and redshift $z_i$ for an iteration. $M_i$ and $z_i$ are used to compute the purity and completeness corrections, which also have associated uncertainties and are again Gaussian sampled. The number density $n_i$ of the candidate is then:

\begin{equation}
n_i = \frac{P_i / C_i}{V}
\end{equation}

where $P_i$ and $C_i$ are the purity and completeness for the given iteration, and $V$ is the comoving volume, which is $9.82 \times 10^6$ Mpc$^3$ for our redshift range of 0.55 to 1.37 and effective transverse survey area of 1.4 square degrees. $n_i$ is added to its mass bin, assigned by $M_i$. As more overdensity candidates are assigned to the same mass bin, the larger the total number density in the bin grows. We repeat this process 1000 times and then take the median of all iterations as the final number densities for each mass bin, with the 16th and 84th percentiles as approximate 1$\sigma$ uncertainties.

\begin{table}
\caption{Overdensity Candidate Parameters}
\label{tab.grid}
\begin{tabular}{ll}
\hline
Parameter & Values \\
\hline
Mass Fit (DETECT\textunderscore THRESH $\sigma$) & Original \citep{hung20}, 4, 5 \\
DEBLEND\textunderscore MINCONT			& 0.03, 0.01, 0.005, 0.003 \\
Linking Radius (Mpc)					& 1.0, 0.50, 0.25 \\	
Using Known Structures					& Yes, No \\
Using $R_{200}$ Correction				& Yes, No \\  [0.7ex]
\hline
\end{tabular}
\begin{flushleft}
Each parameter listed here will affect how many overdensity candidates are found and at what mass, directly affecting whatever mass function we attempt to derive. As we do not know \emph{a priori} the optimal set of parameters to use, we consider reasonable ranges of values for five total parameters, described in \S\ref{sec.finding} and \S\ref{sec.massfuncdata}, giving 144 unique arrangements. 
\end{flushleft}
\end{table}

\begin{figure*}
\centering
\includegraphics[width=2\columnwidth]{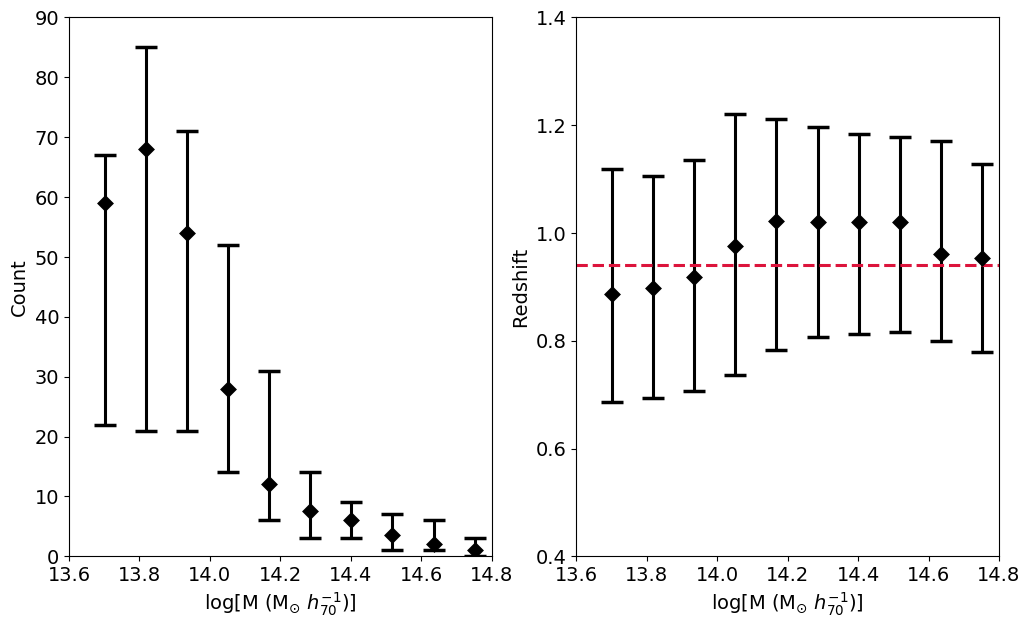}
\caption{Our sample of overdensity candidates will change based on the choice of input parameters in our detection algorithm. Here we show how the total number (left) and redshifts (right) of overdensity candidates vary by averaging over the ranges of values we set in Table \ref{tab.grid}. The points denote the median number in each mass bin, and the error bars show the 16th and 84th percentiles. For simplicity, we assume all individual candidate mass and redshift uncertainties as well as purity and completeness correction uncertainties to be 0 in this plot. The median total number of candidates is 241 in the plotted mass range. The candidate numbers fall just short of 0 for our highest mass bin, and we see the largest spreads in candidate numbers below 14.2 dex. The dashed red line gives the overall mean redshift of the sample at $z=0.94$, which falls within 1$\sigma$ of the redshift range in every mass bin.}
\label{fig.candidates_all}
\end{figure*}

\subsection{Testing for Eddington Bias}
We tested for the presence of \citet{eddington} bias in our sample of overdensity candidates using a toy cluster mass function. We devised a mock cluster mass function and sampled from it a population of observed synthetic clusters. Redshifts and spectroscopic fractions were generated for each synthetic cluster by uniform randomly sampling the full range of these two parameters of the real overdensity candidate sample. Our purity and completeness estimates were unchanged. We assigned every synthetic cluster the same fixed mass uncertainty and tested two cases: 0.05 dex, the typical uncertainty we see in our real overdensity candidate sample, and 0.15 dex, one of the largest uncertainties in the sample. We assigned the synthetic clusters into mass bins with the same Monte-Carlo method described in \S\ref{sec.massbin} to see if we could recover our toy cluster mass function. Though Eddington bias was always clearly present, we found that our number densities only noticeably deviated from the toy mass function when the size of the mass bin was smaller than the clusters' mass uncertainties. Any deviation was negligible otherwise. Given that we use a much larger mass bin of 0.38 dex for our real mass function, we can consider the effects of Eddington bias as small compared to our typical mass error and purity and completeness corrections.

\subsection{Averaging over Parameter Values}
The overdensity candidates in our catalog will change depending on our choice of parameters, and we do not know \emph{a priori} which choice is optimal for building our cluster mass function. However, we can define for each parameter a reasonable range of values from our prior rigorous testing on both real and mock data in \S4 and 5 in \citet{hung20}. With the values we choose below, we were able to recover high fractions of previously known structures with similar redshift and transverse position offsets from their fiducial coordinates. Likewise, our estimated levels of completeness and purity largely fell within a 5\% variation.

We define our grid in Table \ref{tab.grid} by the set of parameters described in \S\ref{sec.finding}, and we plot the variations in our overdensity candidates in number and redshift in Figure \ref{fig.candidates_all}. The mass fit is dependent on the DETECT\textunderscore THRESH $\sigma$ used, as a higher $\sigma$ decreases the sizes of the SExtractor detections. We use new mass fits drawn from using DETECT\textunderscore THRESH values of 4 and 5$\sigma$. Because we have four unique values for DEBLEND\textunderscore MINCONT, the mass fit will slightly differ for each one. In order to obtain a single mass fit for the same DETECT\textunderscore THRESH value, we compute a mass fit for each DEBLEND\textunderscore MINCONT value, leaving us four sets of best-fit terms for the fitting function in Eq. \ref{eq.massfit}. We then obtain an average mass fit by taking the median for each term, and we treat the median absolute deviation as the uncertainty in the term. We also use our original mass fit from \citet{hung20}, which used a DETECT\textunderscore THRESH of 4$\sigma$, as a means of testing another methodological approach divorced from the choice of parameters described in this paper.

We chose a reasonable range of our DEBLEND\textunderscore MINCONT values by examining by eye five pairs of known structures within 0.2 to 3.5 Mpc in the transverse dimensions and $z<\Delta0.02$ in redshift across different ORELSE fields. The deblending becomes finer for smaller values of DEBLEND\textunderscore MINCONT. Dropping the DEBLEND\textunderscore MINCONT too far runs the risk of breaking apart individual overdensities, so we looked for the most conservative value of DEBLEND\textunderscore MINCONT that was able to deblend a given pair of known structures. For four of the pairs, we respectively found DEBLEND\textunderscore MINCONT values of 0.03, 0.01, 0.005, and 0.003. For the remaining fifth pair, we were not able to split the two substructures without breaking the conglomerate detection in SExtractor into more than two components.

We consider linking radii of 1.0, 0.50, and 0.25 Mpc, which we had previously tested in \citet{hung20}. Finally, we examine the effects of including previously known structures or not, which can bias our cluster mass function to high mass overdensities, and using the $R_{200}$ correction, which will shift our larger overdensity candidates to lower masses. 

In total, we consider 144 unique sets of values over five independent parameters for the purposes of our cluster mass function. For each set, we use the Monte-Carlo method described in \S\ref{sec.massbin} to assign the candidates to each mass bin and compute a number density by taking the median over 1000 iterations. We do this for each set of values, meaning we end up with 144 number densities. Because we do not expect any sample of overdensity candidates to be more indicative of reality than another, we then take the median of these number densities to give us our final mass function, with the 1$\sigma$ upper and lower bounds defined by the 84th and 16th percentiles. 

\begin{figure}
\includegraphics[width=\columnwidth]{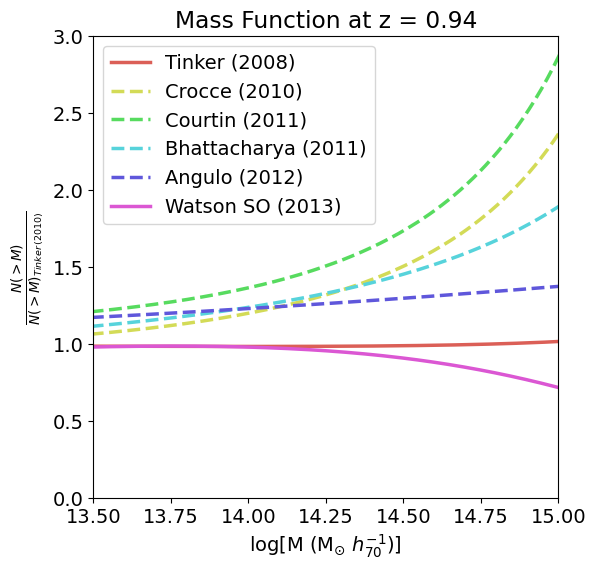}
\caption{We compare the \citet{tinker10} halo mass function with several of the most recent models available in the Halo Mass Function calculator, fixing $z=0.94$, $\Delta = 200$, and using WMAP9 cosmology. Models can either define haloes through the spherical overdensity (SO; solid lines) or Friend-of-Friend (FoF; dashed lines) algorithms. SO halo mass functions are generally taken to be more suited for observational comparisons due to similar spherical definitions of mass, and as expected we see the largest discrepancies with the FoF models, and the discrepancies increase with higher masses. However, all of the plotted models fall within a maximum factor of 3 (average factor of 1.3) of \citet{tinker10} for our mass range of interest, which is smaller than the typical uncertainties we see in our observed number densities.}
\label{fig.hmfs}
\end{figure}

\section{Comparison with Theory}\label{sec.massfunctheory}

For our analysis, we use the halo mass function by \citet{tinker10}, derived from identifying dark matter haloes in N-body simulations of flat $\Lambda$CDM cosmology. Using the spherical overdensity (SO) algorithm, haloes are identified as isolated density peaks. The halo mass is defined in spherical apertures enclosing overdensities $\Delta$, defined as the mean interior density relative to the background. The halo mass function is not the same as the cluster total mass function, but simulations suggest a tight correlation between halo mass and cluster mass proxies \citep[e.g.,][]{kravtsov06, nagai06}. We chose to use the \citet{tinker10} halo mass function as it and \citet{tinker08} are highly cited as a point of comparison for observational studies, and both models are very nearly equal for our redshift and mass ranges. We note, however, that choosing other modern halo mass functions \citep[e.g.,][]{crocce10, courtin11, bhattacharya11, angulo12, watson13} does not significantly change our results as the differences between the models are smaller than our typical number density uncertainties (Fig. \ref{fig.hmfs}).

\begin{figure*}
\centering
\includegraphics[width=2\columnwidth]{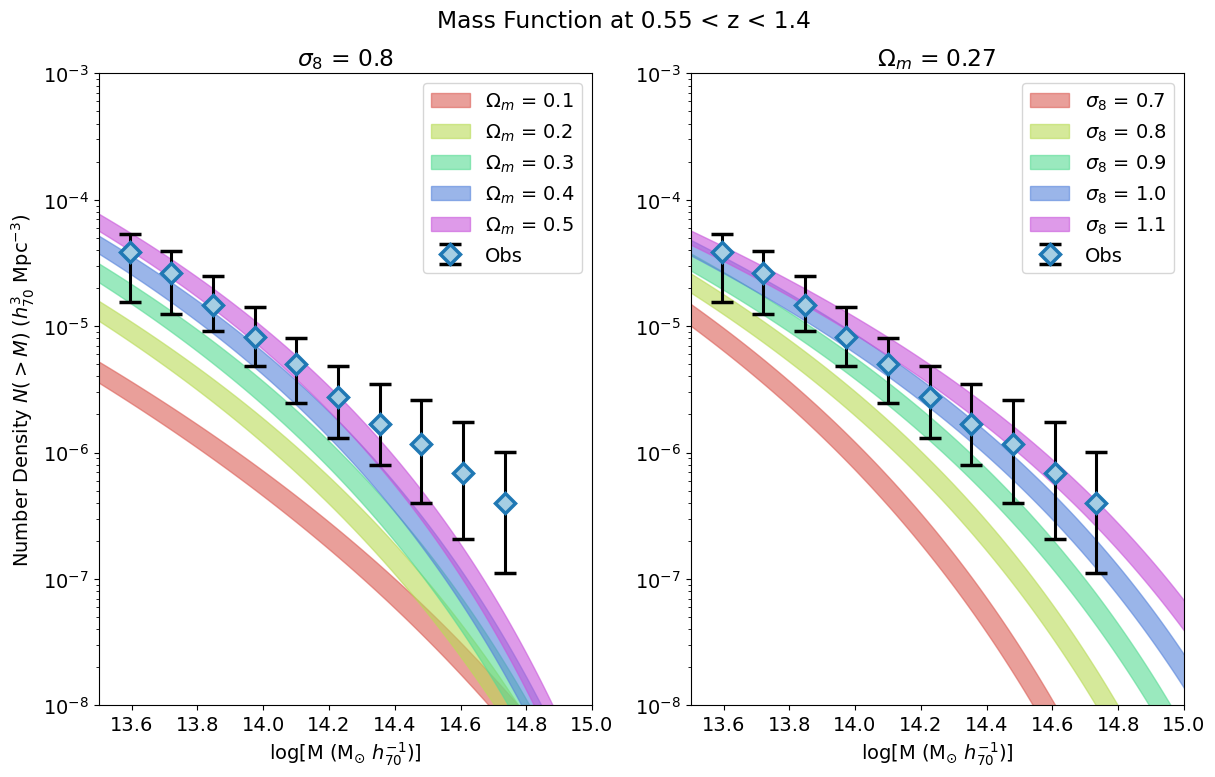}
\caption{We show how the \citet{tinker10} halo mass function changes with $\Omega_{m}$ and $\sigma_{8}$ at a fixed redshift of $z=0.94$, which is the median redshift of our overdensity candidate sample. Each band shows the range covered by $100 < \Delta < 200$. We plot the cluster mass function from our observational data as the median of a total of 144 runs over all parameters described in \S\ref{sec.massfuncdata} and Table \ref{tab.grid}. The lower and upper bounds show the 16th and 84th percentile values.}
\label{fig.massfunc}
\end{figure*}

We generate the theoretical models with the Halo Mass Function calculator \citep[HMFcalc;][]{hmfcalc} available through an online interface as well as a Python package\footnote{Version 3.0.12; \url{https://github.com/steven-murray/hmf}}. The main cosmological parameters that define the halo mass function are $\Omega_{m}$ and $\sigma_{8}$. The other parameters do not strongly affect the halo mass function, so we keep them fixed, adopting the nine-year {\it Wilkinson Microwave Anisotropy Probe} \citep[WMAP9;][]{wmap} parameters, which is available in HMFcalc as a pre-defined cosmology. We plot our observational points against several sets of $\Omega_{m}$ and $\sigma_{8}$ in Figure \ref{fig.massfunc}. We fix the redshift to $z=0.94$, which is the mean redshift of our overdensity candidate sample, a mean which does not depend strongly on structure mass (see Figure \ref{fig.candidates_all}). When comparing $R_{\mathrm{circ}}$ and $R_{200}$ across our sample, we have seen the former be consistently larger. Though we try to correct our masses to roughly $M_{200}$, we plot $\Delta$ ranges of 100 to 200 to compare with larger radii due to our ignorance in how our masses are constructed over the same spatial extents (Fig. \ref{fig.massfunc}).

\begin{figure*}
\centering
\includegraphics[width=2\columnwidth]{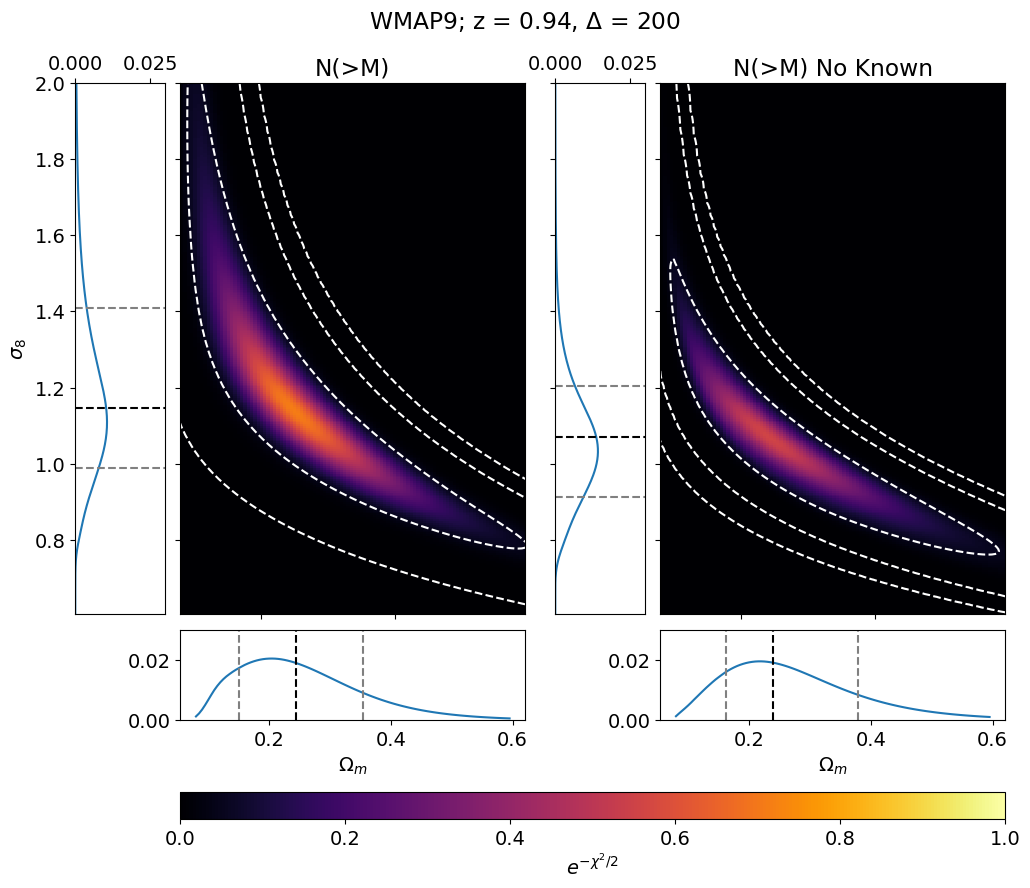}
\caption{We fit for $\Omega_{m}$ and $\sigma_{8}$ using a \citet{tinker10} halo mass function at $z=0.94$, with $\Delta=200$, and using WMAP9 cosmology. The left panel includes all overdensity candidates in our sample, while the right panel excludes all previously known structures. Because of how the ORELSE survey was targeted, the previously known structures are among the largest mass overdensities in our sample. We thus attempt to migitate our higher than predicted number densities in our mass function by excluding them. The white contours in the plot show the 68.3\%, 90\%, and 95.4\% confidence regions. The best-fit values are given by the likelihood maximum in the grid, and the 1$\sigma$ uncertainties are given by the 16th and 84th percentiles of the 1D folded likelihood functions. We show the fitted halo mass functions in Figure \ref{fig.bfit}.}
\label{fig.banana}
\end{figure*}

\begin{figure}
\includegraphics[width=\columnwidth]{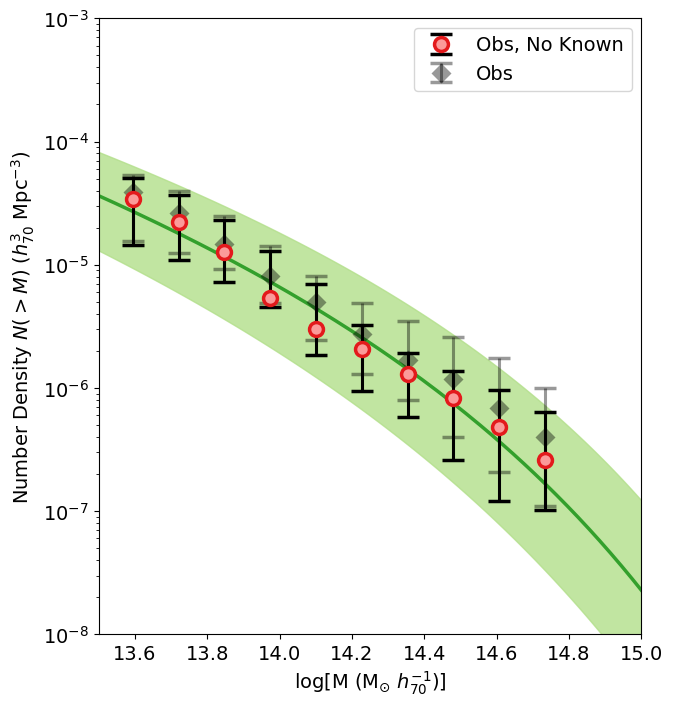}
\caption{We plot the best-fit \citet{tinker10} halo mass functions with the parameters found in Figure \ref{fig.banana}. The solid line follows the best-fit parameters, while the shaded region shows the maximum variation among the 1$\sigma$ ranges for $\Omega_{m}$ and $\sigma_{8}$. }
\label{fig.bfit}
\end{figure}

Our observational points consistently appear high compared to concordance cosmology, which is likely a consequence of our survey being targeted around previously known clusters. Despite this, we can still attempt to fit our observed points for $\Omega_{m}$ and $\sigma_{8}$ to demonstrate the proof of concept that such constraints are possible from $z\sim1$ optical/NIR surveys. In order to constrain $\Omega_{m}$ and $\sigma_{8}$, we define a grid of values in 0.005 steps for $0.080 < \Omega_{m} < 0.600$ and $0.600 < \sigma_{8} < 2.000$ which we iterate over in the \citet{tinker10} halo mass function at $z=0.94$, with $\Delta=200$, and using WMAP9 cosmology. $\Omega_{m}$ is varied in HMFcalc such that the total density parameter $\Omega_{\mathrm{tot}}$ remains flat. At each point in the grid, we measure the $\chi^2$ difference between the cluster and halo mass functions using a standard least squares method, which is transformed to a likelihood by $e^{-\chi^2/2}$. When fitting for $\Omega_{m}$ and $\sigma_{8}$, we split our observed number densities into two groups: one that contains all overdensity candidates as described in \S\ref{sec.massfuncdata} and one excluding all previously known structures (Fig. \ref{fig.banana}). Depending on the choice of SExtractor detection and deblending parameters, we recover between 77 to 93\% of the 56 previously known structures in the ORELSE fields. The known structures constitute 13 to 20\% of all overdensities in our sample by number. At masses greater than $\log (M/M_{\odot}) = 14.5$, however, the known structures make up between 53 to 82\% of the sample. The known structures are among the most massive in our sample, and thus we expect them to bias our observed high-mass densities.

We find $\Omega_{m} = 0.250^{+0.104}_{-0.099}$ and $\sigma_{8} = 1.150^{+0.260}_{-0.163}$ among our complete sample, and $\Omega_{m,nk} = 0.240^{+0.139}_{-0.077}$ and $\sigma_{8,nk} = 1.070^{+0.133}_{-0.157}$ when the known structures are removed. We plot these fits against their respective observed points in Figure \ref{fig.bfit}. Under the $\Lambda$CDM model, WMAP9 gives $\Omega_{m} = 0.279 \pm 0.025$ and $\sigma_{8} = 0.821 \pm 0.023$\footnote{\url{https://lambda.gsfc.nasa.gov/product/map/dr5/params/lcdm_wmap9.cfm}}. Our best-fit $\Omega_{m}$ values agree with the concordance value within 1$\sigma$, while $\sigma_{8}$ is discrepant at the $\sim$2 and $\sim$1.5$\sigma$ levels when the known structures are included and excluded, respectively. From the right panel in Figure \ref{fig.massfunc}, we see that our observed points closely follow a line of fixed $\sigma_{8}$ for $\Omega_{m} = 0.27$, which consequently shifts $\sigma_{8}$ higher to compensate. Likewise, the best-fit $\Omega_{m}$ of our sample with and without the known structures are very similar, but $\sigma_{8}$ is slightly smaller for the sample without the known structures. However, we note that a considerable fraction ($\sim$50\%) of our high-mass overdensity candidates are very close in redshift/transverse space to known massive systems. Even with the known structures removed from the sample, the presence of these close candidates grants substantial power to the high end of the mass function, which in turn elevates the $\sigma_{8}$ parameter.

\section{Discussion}\label{sec.discuss}

We present our $\Omega_{m}$ and $\sigma_{8}$ fits as a proof of concept that cosmological fitting can be done with optical/NIR data at $z\sim1$, which to the best of our knowledge has not been done before outside of the relatively local universe. As a result of our consistently high number densities, especially at the high mass end (see discussion in \S\ref{sec.prev} and \ref{sec.massfunctheory}), while our best-fit $\Omega_{m}$ is consistent within $1\sigma$ with the concordance value, our best-fit $\sigma_{8}$ is roughly $2\sigma$ higher than the equivalent concordance value. However, the ORELSE survey was by design targeted around known large-scale structures, so we would expect to see more galaxy overdensities per volume than an equivalent field survey. Though other recent studies such as \citet{abbott20} have found a tension in their derived cosmological parameters due to disagreements between different mass proxies, we do not share similar concerns, at least at the high-mass end, since our dynamical mass estimates are within the error bars of the X-ray, lensing, and SZ mass measurements found in other studies. However, the issue in \citet{abbott20} was primarily at the low-mass end. To check if this is potentially an issue for our results, we excluded the two lowest mass bins in our observational mass function and re-derived the cosmological parameters. We found no meaningful difference in our results, with $\Omega_{m}$ and $\sigma_{8}$ being entirely within the error bars of the values we found in \S\ref{sec.massfunctheory}.

Optical-wavelength mass function studies conducted for the local universe have recently begun to emerge. \citet{abdullah20} derived a cluster mass function using 756 Sloan Digital Sky Survey \citep[SDSS;][]{sdss}  clusters with masses estimated from the virial theorem. Their sample had a mean redshift of $z = 0.085$ and a similar mass range to our work. The authors find good agreement with the \citet{tinker08} model, only significantly falling short at $\log (M/M_{\odot}) < 14$, suggesting a possible sample incompleteness which we were able to avoid at the same mass threshold. However, due to their smaller density uncertainties across the mass range, \citet{abdullah20} were able to recover tighter cosmological constraints of $\Omega_{m} = 0.310^{+0.023}_{-0.027}$ and $\sigma_{8} = 0.810^{+0.031}_{-0.036}$, with systematic errors of $\pm0.041$ and $\pm0.035$ respectively. Though our sample is at an order of magnitude higher redshift, our errors are only two to three times as large as the combined random and systematic errors found by \citet{abdullah20}.

Cluster count studies at higher redshifts have traditionally only been done with X-ray and SZ surveys, though even then mass function studies have been few. \citet{vikhlinin09} derived a mass function with two cluster samples. The high-redshift sample had 37 clusters derived from the 400 square degree ROSAT serendipitous survey \citep{rosat} and covered the redshift range $0.35 \le z \le 0.90$. The low-redshift sample consisted of the 49 highest flux clusters detected in the ROSAT All-Sky Survey and was over $0.025 \le z \le 0.25$. Both samples were later observed by the Chandra X-ray Observatory, providing spectral data that enabled several high-quality total mass estimators. Cluster masses are estimated using the X-ray luminosity and total mass relation. Both samples approximately cover the mass range $14 < \log (M/M_{\odot}) < 15$. With the \citet{tinker08} halo mass function, the authors find $\Omega_{m} = 0.255\pm0.043$ and $\sigma_{8} = 0.813\pm0.013$, with respective systematic errors of $\pm0.037$ and $\pm0.024$. However, the authors find that the constraints on $\sigma_{8}$ do not significantly change when measured with only the low-redshift sample and then again with the total sample including the high-redshift data, which the authors argue implies the $\sigma_{8}$ measurement is dominated by the more accurate local cluster data.

\citet{bocquet19} derived cosmological constraints with a galaxy cluster sample of 365 candidates over the redshift range $0.25<z<1.75$ from the 2500 square degree SPT-SZ survey. Some clusters in the sample were also supplemented with optical weak gravitational lensing or X-ray measurements. Through using SZ, X-ray, and weak lensing mass proxies, the sample is estimated to cover a mass range of approximately $14.4 < \log (M/M_{\odot}) < 15.3$. The authors find constraints of $\Omega_{m}= 0.276\pm0.047$, $\sigma_{8}= 0.781\pm0.037$ with the \citet{tinker08} halo mass function.

With ORELSE and VMC mapping, we have the advantage of being sensitive to lower mass ranges than traditional X-ray and SZ survey studies. X-ray studies, however, will soon enjoy a boon of data with the ongoing all-sky survey by the extended Roentgen Survey with an Imaging Telescope Array \citep[eROSITA;][]{erosita} instrument on the Spectrum-Roentgen-Gamma (SRG) mission, which will produce on the order of 10,000 detections of the hot intergalactic medium of galaxy clusters. VMC mapping itself is adaptable to any similar photometric and spectroscopic dataset, and thus has great potential when combined with future, larger optical surveys.

Spectroscopic redshifts are tremendously useful for cluster studies as they provide highly accurate information on where galaxies are distributed along the line-of-sight, but they have been traditionally difficult to obtain due to their large time commitment. The Subaru Prime Focus Spectrograph \citep[PFS\footnote{\url{https://pfs.ipmu.jp/intro.html}};][]{pfs} is an optical and NIR wavelength spectrograph expected to be ready for scientific use in 2022. Situated on the 8.2-m Subaru Telescope, PFS is capable of obtaining spectra of galaxies that were technologically out of reach before. With a 1.3 degree diameter field-of-view, it is capable of simultaneous spectral observation of up to 2400 targets. The forthcoming 100 night PFS cosmology survey aims to sample galaxies over a redshift range of $0.8 \le z \le 2.4$ and a comoving volume of 9$h^{-3}$ Gpc$^{3}$, approximately a thousand times larger than ORELSE's spectroscopic footprint. The ground-based Maunakea Spectroscopic Explorer\footnote{\url{https://mse.cfht.hawaii.edu/}} is a 11.25-m telescope that will replace the 3.6-m Canada-France-Hawaii Telescope (CFHT). Construction on the telescope is anticipated to begin in 2023, with full science operations commencing in August 2026. Its spectrographs can accomodate roughly 3,000 spectra simultaneously. Combined with the telescope's 1.5 square degree field-of-view, it will be able to obtain many more high-quality spectroscopic redshifts from the ground with less time than was possible before.

Photometric redshifts are less accurate than spectroscopic redshifts, but they generally have more uniform spatial distributions and thus enable more complete mapping of the density field of galaxies when combined with spectroscopic redshifts. Photometric redshifts complete to deeper magnitudes will be in no short supply with upcoming all-sky surveys. The ground-based Large Synoptic Survey Telescope \citep[LSST;][]{lsst} is an optical survey expected to begin operations by 2022, with the aim of uniformly observing 18,000 square degrees of the sky 800 times over 10 years. Its six-band photometry will yield photometric redshifts for billions of galaxies. The European Space Agency mission Euclid\footnote{\url{https://sci.esa.int/web/euclid/}} is a space telescope operating at optical and NIR wavelengths planned to launch in 2022. It will measure the redshifts of galaxies out to $z\sim2$ over its nominal six-year mission. Its wide survey component will cover 15,000 square degrees of sky. Its deep survey will reach two magnitudes deeper in three fields with an area totaling 40 square degrees.

A methodology with high purity and completeness such as VMC mapping will be able to take full advantage of this wealth of high-quality data and yield promising results for optical/NIR cluster cosmology in the decades to come.

\section{Summary and Conclusion}\label{sec.conclusion}

With the extensive photometric and spectroscopic dataset from the ORELSE survey and Voronoi tessellation Monte-Carlo mapping, we have derived the first observational cluster mass function at optical and NIR wavelengths outside of the relatively local universe. 

Our original methodolgy in \citet{hung20} recovered 51 previously known structures and found 402 new overdensity candidates over the redshift range $0.55<z<1.37$ and mass range $10.2<\log(M/M_{\odot})<14.8$.  However, we had for the most part set aside the issue of separating blended structures in favor of the most general case of finding any overdensity in the data. As the cluster mass function reports the number density as a function of mass, we needed in this paper to take caution with what candidates in our sample were single structures or not. In total, we had five independent parameters that affected the numbers and masses of candidates we obtained from the same dataset. We also limited our sample to the mass range $13.6<\log(M/M_{\odot})<14.8$, which is where we had purity and completeness estimates from our tests with mock catalogs. We had 144 unique sets of values for the five overdensity candidate parameters, where the median total number of overdensity candidates was 241 and the median redshift was $z=0.94$. We derived the cluster mass function through treating our overdensity candidates sample with a Monte-Carlo scheme and applied purity and completeness corrections as functions of redshift, mass, and spectroscopic fraction.

We compared our observational mass function to the \citet{tinker10} halo mass function, set to $z=0.94$ to match the median redshift of our sample, and using $\Delta=200$ and WMAP9 cosmology. We find cosmological constraints of $\Omega_{m} = 0.250^{+0.104}_{-0.099}$ and $\sigma_{8} = 1.150^{+0.260}_{-0.163}$. While our $\Omega_{m}$ value agrees with the concordance value within 1$\sigma$, our $\sigma_{8}$ value is high by approximately 2$\sigma$. This discrepancy is a consequence of our inflated observed number densities, brought about because ORELSE was designed to be targeted at known large-scale structures. In an attempt to mitigate this, we fitted for $\Omega_{m}$ and $\sigma_{8}$ again after removing all previously known structures from our sample, which gave us constraints of $\Omega_{m,nk} = 0.240^{+0.139}_{-0.077}$ and $\sigma_{8,nk} = 1.070^{+0.133}_{-0.157}$, dropping the discrepancy in $\sigma_{8}$ to roughly 1.5$\sigma$.

The $\Omega_{m}$ and $\sigma_{8}$ constraints we present here are meant to be taken as a proof of concept that pure optical/NIR cluster abundance can be a viable cosmological probe at moderately high redshifts. Though it has limitations when applied to data obtained through a biased survey strategy, our methology has strong potential when combined with the several large optical surveys on the horizon, which will yield many more photometric and spectroscopic redshifts than what was possible to obtain before. Along with advancements in X-ray surveys which will offer complementary results for investigating cluster evolution, we can expect cluster-based constraints to grow into an even more powerful cosmological probe in the near future.

\section*{Acknowledgements}

{\footnotesize
We would like to thank Steven Murray and Chris Power for their prompt guidance to our questions with using HMFcalc. We would also like to thank Alexey Vikhlinin for his useful comments on our mass function analysis and comparisons. We also thank the anonymous referee for their valuable suggestions. Some of the material presented in this paper is supported by the National Science Foundation under Grant Nos. 1411943 and 1908422. This work was additionally supported by the France-Berkeley Fund, a joint venture between UC Berkeley, UC Davis, and le Centre National de la Recherche Scientifique de France promoting lasting institutional and intellectual cooperation between France and the United States. This study is based, in part, on data collected at the Subaru Telescope and obtained from the SMOKA, which is operated by the Astronomy Data Center, National Astronomical Observatory of Japan. This work is based, in part, on observations made with the Spitzer Space Telescope, which is operated by the Jet Propulsion Laboratory, California Institute of Technology under a contract with NASA. UKIRT is supported by NASA and operated under an agreement among the University of Hawaii, the University of Arizona, and Lockheed Martin Advanced Technology Center; operations are enabled through the cooperation of the East Asian Observatory. When the data reported here were acquired, UKIRT was operated by the Joint Astronomy Centre on behalf of the Science and Technology Facilities Council of the U.K. This study is also based, in part, on observations obtained with WIRCam, a joint project of CFHT, Taiwan, Korea, Canada, France, and the Canada-France-Hawaii Telescope which is operated by the National Research Council (NRC) of Canada, the Institut National des Sciences de l'Univers of the Centre National de la Recherche Scientifique of France, and the University of Hawai'i. Some portion of the spectrographic data presented herein was based on observations obtained with the European Southern Observatory Very Large Telescope, Paranal, Chile, under Large Programs 070.A-9007 and 177.A-0837. The remainder of the spectrographic data presented herein were obtained at the W.M. Keck Observatory, which is operated as a scientific partnership among the California Institute of Technology, the University of California, and the National Aeronautics and Space Administration. The Observatory was made possible by the generous financial support of the W.M. Keck Foundation. We thank the indigenous Hawaiian community for allowing us to be guests on their sacred mountain, a privilege, without which, this work would not have been possible. We are most fortunate to be able to conduct observations from this site.}

\section*{Data availability}

{\footnotesize
The data underlying this article will be shared on reasonable request to the corresponding author.}

\bibliographystyle{mnras}
\bibliography{massfunc} 

\bsp	
\label{lastpage}
\end{document}